\documentclass[10pt,aps,prb,twocolumn]{revtex4-2}
\usepackage[english]{babel}
\usepackage[T1]{fontenc} %newley package
\usepackage[utf8]{inputenc} 
\usepackage{verbatim}
\usepackage{graphicx}
\usepackage{graphics}
\usepackage{color}
\usepackage{textcomp}
\usepackage{amsfonts}
\usepackage{amsmath}
\usepackage{hyperref}
\usepackage{amssymb}
\usepackage{lipsum}
\usepackage{mathtools}
\usepackage{tabularx}
\usepackage{subcaption}
\newcolumntype{Y}{>{\centering\arraybackslash}X}
\usepackage{times}
\begin{document}

%\title{Pair heterogeneous mean field approximation for outbreaks in the
%asynchronous susceptible-infected-removed model in complex networks}
%
\title{Estimating thresholds for asynchronous susceptible-infected-removed model on complex networks}

\author{D.S.M. Alencar }
\email{davidalencar@ufpi.edu.br}
\affiliation{Departamento de F\'{\i}sica, Universidade Federal do Piau\'{i}, 57072-970, Teresina - PI, Brazil}
\author{T.F.A. Alves}
\affiliation{Departamento de F\'{\i}sica, Universidade Federal do Piau\'{i}, 57072-970, Teresina - PI, Brazil}
\author{F.W.S. Lima}
\affiliation{Departamento de F\'{\i}sica, Universidade Federal do Piau\'{i}, 57072-970, Teresina - PI, Brazil}
\author{R.S. Ferreira}
\affiliation{Departamento de Ciências Exatas e Aplicadas,
Universidade Federal de Ouro Preto,
35931-008, João Monlevade - MG, Brazil}
\author{G.A. Alves}
\affiliation{Departamento de Física, Universidade Estadual do Piauí, 64002-150, Teresina - PI, Brazil}
\author{A. Macedo-Filho}
\affiliation{Departamento de Física, Universidade Estadual do Piauí, 64002-150, Teresina - PI, Brazil}

\date{Received: date / Revised version: date}

\begin{abstract}
We use the pair heterogeneous mean-field (PHMF) approximation for an asynchronous version of the susceptible-infected-removed (SIR) model to estimate the epidemic thresholds on complex quenched networks. Our results indicate an improvement compared to the heuristic heterogeneous mean-field theory developed for one vertex (HMF) when the dynamic evolves on top random regular and power-law networks. However, there is a slight overestimation of the transition point for the later network type. We also analyze scaling for random regular networks near the thresholds. For this region, collapses were shown at the subcritical and supercritical phases.
\end{abstract}


\pacs{AA}

\maketitle

\section{Introduction} \label{sec:outline}

The Susceptible-Infected-Removed (SIR) model gained visibility over the years because it is a simple model for producing permanent immunity in a community or population. The model separates individuals into three types of compartments or states: susceptible (S), infected (I), and removed (R) \cite{Keeling-2007}. 
Susceptible individuals can become infected through contact with infected neighbors and infected individuals, who, in turn, can be infected.
Recovered and transferred to the removed compartment.

 A simple way to model an epidemic spreading is by applying it in a network, where it is a manner to express human contacts and quantify the strength of this last. The first equations of the temporal evolution for the three compartments are associated with the well-stirred approximation, where all individuals interacted with each other \cite{kermackI, kermackII, kermackIII}.

  The SIR model has been extensively studied in complex networks and lattices \cite{sir-seeds, alencar2022epidemic, RPS-review, Santos-2020}. Standard epidemic models define a node from a network as an individual in one of three states $S$, $I$ or $R$. In a set of $N$ nodes, the adjacency matrix $\textbf{A}_{N\times N}$ with entries $A_{ij}$ quantifies interaction between two nodes $i$ and $j$. For simple and undirected networks, $A_{ij}$ takes the value equal to $1$ if there is a connection; otherwise, it is $0$. The references \cite{alencar2022epidemic, Santos-2020, Tome_2011, Tome-2010} explored the capability of the model to exhibit phase transitions accompanied by exponents and thresholds (transition points) on these complex networks and regular lattices too. The essential parameters to see phase transitions are infection and recovery rate, $\mu_{c}$ and $\mu_{r}$, respectively, or their ratio, known as reproduction number, $\lambda = \mu_{c}/\mu_{r}$. The transition is theoretically previewed by simplifying state correlations and adjacency matrix spectral properties, called quenched mean-field (QMF) approximation \cite{RPS-review}. Another approach is to use network degree distribution $P(k)$, considering nodes with the same degree as statistically equivalent and change $\textbf{A}$, therefore, by the conditional linking probability of nodes with degree $k$ being connected with other nodes with $k'$ as $P(k'|k)$. This procedure is named heterogeneous mean-field (HMF)\cite{RPS-review, barrat2008, newman-book}. 
% On simple graphs, undirected, without self-connections, weighted or double connections between nodes, this structure is described by  adjacency matrix $A_{ij}$: is $1$ if $i$ connects $j$ (and vice-versa) and $0$ otherwise.  Standard epidemic models defines a site as a individual and the $A_{ij}$ is the relationship between others individuals of network of size $N$.\cite{Tome-2010}\cite{Stauffer-1992}\cite{ALENCAR2020}.
% For almost any type of network, the model presents phase transition accompanied by exponents and transition points. Complex networks generally present small-world phenomenon, which is why we can approximate the correlations between states by their products (law of mass action).

Recently, we report, analytically via HMF and by simulations on a modified version that reduces the number of possible contacts per time unit, named asynchronous SIR model\cite{Tome-2010}, that it presents continuous phase transition at long times with removed nodes density $r_{\infty}$ being the order parameter and $\lambda$ controlling it. An analog for the model is the usual contact process (CP) model \cite{Tome-2015}. At transition point, $\lambda_{c}$, order parameter scales with network size $N$ accompanied by corrections as $r_{\infty}\sim (gN)^{-1/2}$ where $g$ is\cite{boguna-2004}
\begin{equation}
g = \langle k^{2}\rangle/{\langle k \rangle}^{2}.
\label{momentratio}
\end{equation}

 This factor can make the transition non-universal with mean-field (MF) exponents depending on network connectivity distribution $P(k)$. When the model is treated on power-law networks (PLN), where the probability of finding a  node with degree $k$ is $P(k) \sim k^{-\gamma}$ for $\gamma \geq 2$, and the ratio $g$ is diverging with size $N$ for $2<\gamma\leq 3$, forming a high heterogeneous network, is possible to see a change on ordinary mean-field exponent $1/2$. In the reference \cite{alencar2022epidemic}, some issues, such as the scaling concordance and dispersion at the low and high prevalence of removed nodes, respectively, were shown for PLN (various $\gamma$) and Barabási-Albert network \cite{barabasi1999}, which is a weakly correlated network \cite{RPS-2005}. 
  
In a broader sense, mean-field approximations can be improved when we draw equations for correlations, as previous references have already reported for other epidemic models \cite{daMata_2014, Mata_2013, multiplex}. The statistical equivalence can be treated to neighboring pairs. In this work, we consider pair heterogeneous mean-field (PHMF), already used to extensively studied contact process model \cite{daMata_2014}, to find more convergence between simulated and theoretical prediction thresholds for the asynchronous SIR model. We will show results for the random regular network (RRN) and PLN. Simulations show an agreement for the first network kind and precise but slightly overestimated results encountered for the second. RRNs' results will also show that the scaling form calculated works well on the subcritical phase. 

This work is organized as follows: in section \ref{asynchronous}, we revisit the model at the HMF level. In section \ref{pair} is devoted to heterogeneous pair approximation (PHMF). In section \ref{results}, we show simulation results and discussion of the model, and in section \ref{conclusion}, we present our conclusions.   %I believe leaving the sections in separate files is more organized, change it if you desire 
\section{Asynchronous SIR Model } 
\label{asynchronous}
\begin{comment}
Route to write this 
\begin{itemize}
    \item Cite briefly droplet theory to access FSS 
    \item Show it semi-adhoc threshold $\lambda_{c} = \langle k \rangle /(\langle k \rangle - 1)$
    \item Show Pair-Heterogeneous and it transcendental eq. to got $\lambda_{c}$
    \item Show if transcendent equation reaches actual values. 
    \item Give to Sylvie's article the contribution for it find similar equations
\end{itemize}
\end{comment}

\begin{comment}
 (Susceptible-Infected-Removed) model is one that is able to assemble a temporal evolution of a disease with  permanent immunity in a population \cite{Keeling-2007}. Each individual can be in only one of three states: susceptible (healthy), infected, or removed. The infection process is simple: an healthy individual takes infection by a contact, with $\mu_{c}$ rate, with a another infected one, or, if the first is infected, cures it spontaneously at rate $\mu_{r}$. 
 
   In a network, the relations between theses individuals or node is quantified by adjacency matrix $A_{ij}$ \cite{newman-book}. For simple, undirected, unweighted network and single component network, if there is connection between node $i$ and $j$, $A_{ij}$ takes the value equal $1$, otherwise, is $0$. The number of possible contacts is equal node degree $k_{i} = \sum_{i}A_{ij}$.
\end{comment}
To couple the SIR dynamics with a network, a node can be in one of these three states (S, I, and R). A variable $\epsilon_{i}$ is the state of node $i$, with values $0, 1, 2$ for susceptible, infected, and removed, respectively. The S, I, and R states are denoted by $s_{i} = [0_{i}]$, $\rho_{i} = [1_{i}]$ and $r_{i} = [2_{i}]$, respectively.  The dynamics follow this scheme: an infection event ($S \rightarrow I$) on susceptible node $i$ by its infected neighbor $j$ occurs with the rate $\mu_{c}/k_{j}$; or if the node is infected, a spontaneous recover ($I \rightarrow R$) with rate $\mu_{r}$ recover it. The term $1/k_{j}$ gives an equal chance of infection for any neighbor of $j$. This inclusion contrasts with many studies of the standard SIR model, from which it is disregarded ($1/k_{j} \rightarrow 1 $). The main difference between the two approaches is the number of possible contacts per unit of time in the neighborhood: the first allows only one, and the last, $k_{j}$ contacts. Some works include the term used here on regular and disordered lattice simulations \cite{tome2010critical, Tome_2011, Tome-2015, ALENCAR2020, santos2020epidemic}, where the dynamics were called asynchronous SIR model, to analyze the epidemic spreading starting from a single seed.

For any time, the three densities of individuals on an 'i' node have a constraint:
\begin{equation}
      s_{i}(t) + \rho_{i}(t) + r_{i}(t) = 1.
      \label{constraint}
\end{equation}

\begin{figure}[!t]
     \centering
     \includegraphics[scale = 0.3]{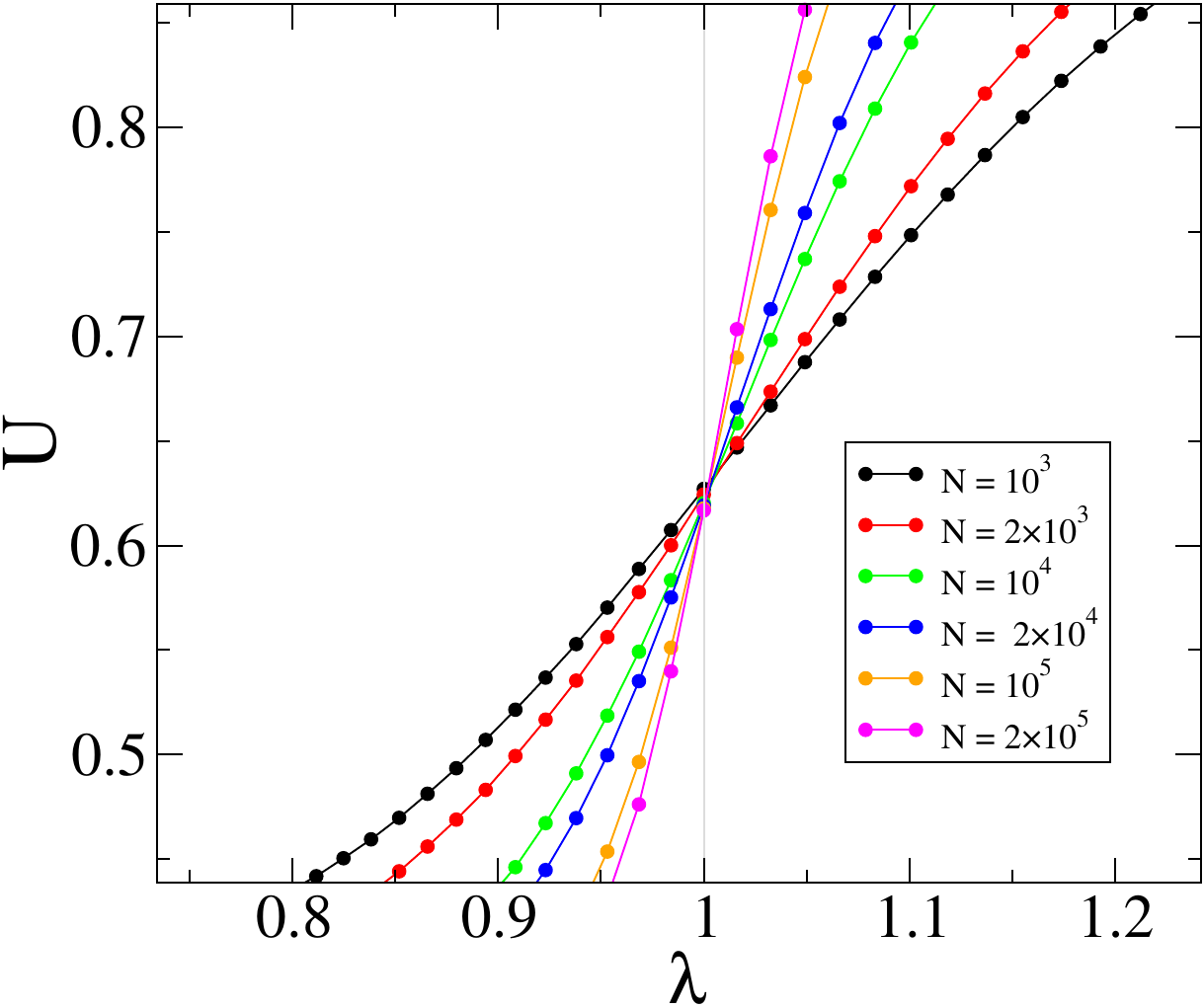}
     \caption{(Color online). Third order cumulant from eq. \ref{cumulant} for annealed PLNs with $N$ nodes with degree exponent $\gamma = 2.50$. Crossings for increasing $N$ towards the transition point from theory $\lambda_{c} = 1$ (vertical line).}
     \label{cumulant-annealed}
\end{figure}
 %referencia texto ronan%  
The adjacency matrix is sparse for many real and synthetic networks. In addition, the networks share the small-world property \cite{barabasi-review}. The small-world property induces shortcuts between network nodes, which decreases the average distances logarithmically or slower than its size. The effect on the model is an approximation for correlations involving susceptible and infected neighbors. This procedure gives only a threshold for the model based on the spectral property of $A_{ij}$, and for the present model, it always results in $\mu_{c}/\mu_{r} = 1$ as the threshold, and it is called quenched mean-field (QMF).

The mean-field hypothesis can be generalized for heterogeneous and homogeneous networks when the adjacency matrix has full information about the network. We can approach the quenched disorder by ensemble averages, where the probability of a node with degree $k$ connecting with a node of degree $k'$ is proportional to $kP(k'|k)$ \cite{barrat2008, RevModPhys.80.1275}. $P(k'|k)$ is the conditional probability of a $k$-degree node connecting a $k'$-degree node. In this way, we can treat network nodes in a heterogeneous manner, considering nodes with the same degree governed by the same equations, which is the heterogeneous mean-field (HMF) approximation. The concentration variables change their label (e.g., $\rho_{i} \equiv \rho_{k} = [1_{k}]$, and similar for the other two) and interact among other nodes mediated by $kP(k'|k)$. For the asynchronous SIR model, one can construct the time evolution equations by mass action law for $k$-degree nodes on generalized heterogeneous form 
\begin{equation}
    \left\lbrace
    \begin{split}
    & \dfrac{ds_{k}}{dt} = -\mu_{c} k\sum_{k'}\frac{P(k'|k)\phi_{kk'}}{k'} \\
    & \dfrac{d\rho_{k}}{dt} =\mu_{c} k\sum_{k'}\frac{P(k'|k)\phi_{kk'}}{k'} - \mu_{r} \rho_{k}(t), \\
    & \dfrac{dr_{k}}{dt} = \mu_{r}\rho_{k}.
    \label{sirk}
    \end{split}
    \right.
\end{equation}
   
The first term in eq. \ref{sirk} is the probability of generating an infection on a $k$-degree node, which depends on the probability of finding pairs $\phi_{kk'}$ of susceptible and infected sites of degree $k$ and $k'$, respectively, on the probability of a connection between these nodes, given by $kP(k'|k)$, and on the rate $\mu_{c}/k'$ of successful contamination. The second term represents spontaneous recovery with rate $\mu_{r}$. The HMF procedure treats $\phi_{kk'}$ approximately independently. Thus, $\phi_{kk'} \approx s_{k}\rho_{k'}$ and this rewrites previous equations as 
   \begin{equation}
    \left\lbrace
    \begin{split}
       & \dfrac{ds_{k}}{dt} = -\mu_{c} s_{k}(t)k\Theta(t), \\
       & \dfrac{d\rho_{k}}{dt} =\mu_{c} s_{k}(t)k\Theta(t) - \mu_{r} \rho_{k}(t), \\
       & \dfrac{dr_{k}}{dt} = \mu_{r} \rho_{k},
    \end{split}
    \right.
    \label{sirkmf}
\end{equation}
where
\begin{equation}
\Theta(t) = \sum_{k'}\frac{P(k'|k)\rho_{k'}}{k'}.
\label{theta}
\end{equation}

The removed concentration are calculated by averaging on $P(k)$, \textit{e.g.}
\begin{equation}
r(t) = \sum_{k}P(k)r_{k}.
\end{equation}
On the uncorrelated networks, the conditional probability is 
\begin{equation}
P(k'|k)= k'P(k')/\langle k \rangle,
\label{conditionalprobability}
\end{equation}
where $\langle k \rangle $ is the mean degree.  For quenched networks, where the connections are fixed on time, a supposition to improve model accuracy works with the possibility of removing one of the edges to contamination from infected nodes, as they cannot transmit the disease to the one that contaminated it \cite{RPS-review, RevModPhys.80.1275}. Thus, an \textit{ad hoc} expression which excludes the source of infection with degree $k'$ is
 \begin{equation} 
    P(k'|k) = \frac{(k'-1)P(k')}{\langle k \rangle}.
    \label{sourceremoval}
 \end{equation}   
Expression \ref{sourceremoval} is a heuristic approximation: for more accuracy in the model, considering the infection source removal, $P(k'|k)$ would need to be re-calculated every time. For equation \ref{theta}, using eq. \ref{sourceremoval} we have
\begin{equation}
 \Theta(t) = \frac{1}{\langle k \rangle }\sum_{k'}\frac{(k'-1)P(k)\rho_{k'}}{k'}.
\label{source}
\end{equation}

 Recently, we proposed a time evolution for HMF equations \ref{sirkmf} and \ref{theta} based on initial conditions from one infection seed \cite{alencar2022epidemic}. The concentrations at $t=0$ are $s_{k}(0) = 1-h$, $\rho_{k}(0) = h$, and $r_{k}(0) = 0$, where $h = 1/N$ represents a single infected individual on entire network. Based on this argument, it was possible to find a transition point $\lambda_{c}$ based on linear stability analysis %\footnote[1]{}
 \begin{equation}
     \lambda^{\text{HMF}}_{c} = \frac{\langle k \rangle}{\langle k \rangle -1}.
     \label{hmfthrehsolds}
 \end{equation}

The threshold on eq. \ref{hmfthrehsolds} has sense only quenched networks (fixed structure). The HMF is expected to be exact in one type of network named annealed, where all connections are rewired between two dynamical steps. Two node correlations are described by their concentration product, and no source removal is needed; therefore, the conditional probability is given by eq. \ref{conditionalprobability}, and $\lambda_{c} = 1$ always, independent of $P(k)$ shape. We check this threshold in Figure \ref{cumulant-annealed} using an annealed network with degree distributed as power-law $P(k) \sim k^{-\gamma}$.

  The density of removed nodes at steady state, $r_{\infty}$ was obtained based on initial conditions, and it scales approximately as 
\begin{equation}
     r_{\infty} - h \approx \left(  \frac{gN}{2a}\right)^{-1/2}F_{N}\left[\left(   \frac{1}{\lambda_{c}}- \frac{1}{\lambda} \right) \left(  \frac{2N}{ga}\right)^{1/2}\right],
     \label{fss-sir}
\end{equation}
where 
\begin{equation}
    a = 1 - \langle 1/k \rangle \approx 1,
\end{equation}
$F_{N}$ is a scaling function, and $g$ is given in equation \ref{momentratio}, and represents degree fluctuations of the network. The $g$ scales with the network size $N$ for heterogeneous graphs. Especially for uncorrelated PLNs, using the continuous media approximation \cite{alencar2023droplet, boguna-2004}, we have
\begin{equation}
    g \sim \left\lbrace
    \begin{split}
        & N^{(3-\gamma)/\omega} \text{ for } 2<\gamma<3, \\
        & \ln {N} \text{ for } \gamma=3, \\
        & \text{constant for } \gamma>3,
    \end{split}
    \right.
\end{equation}
where $\omega = [2, \infty]$ is a cutoff exponent.

To improve the mean-field picture, we can track time evolution for $\phi_{kk'}$ pairs a to give a more accurate threshold estimation. The equations are made in the same manner as \ref{sirk}, considering interactions among other pairs of different degree combinations.
\section{ Pair heterogeneous mean-field approximation} \label{pair}

  The HMF theory is inaccurate in estimating thresholds for quenched networks because it approximates correlations evolving the first neighborhood $\phi_{kk'} \approx s_{k}\rho_{k}$. Equations for the time evolution of the first neighborhood, obviously considering only connected nodes, are expected to show an improvement in the estimated threshold. In this way, consider the pairs of nodes in the first neighborhood of the network and approach those in the second, together with network distribution $P(k)$, the statistical equivalence present in the last section is adapted: if a pair of nodes have $k$-degree $k'$-degree, respectively, the probability of they assume a joint state or their concentration is represented by $[\sigma_{k},\sigma_{k'}]$ \cite{diogo-2022, daMata_2014, multiplex}. This approach is named pair heterogeneous mean-field (PHMF).
 
 The possible combinations of these joint states are 
\begin{align}
    \theta_{kk'}&=[2_{k},1_{k'}], & \overline{\theta}_{kk'}&=[1_{k},2_{k'}], & \nu_{kk'}&=[2_{k},2_{k'}], \nonumber \\
    \chi_{kk'}&=[2_{k},0_{k'}], & \overline{\chi}_{kk'}&=[0_{k},2_{k'}], & \psi_{kk'}&=[1_{k},1_{k'}], \nonumber \\
    \phi_{kk'}&=[0_{k},1_{k'}], & \overline{\phi}_{kk'}&=[1_{k},0_{k'}], & \omega_{kk'}&=[0_{k},0_{k'}].
\end{align}

 This implies the following relations
\begin{align}
    \psi_{kk'}   &= \psi_{k'k}    & \phi_{kk'}   &= \overline{\phi}_{k'k},   \nonumber \\
    \omega_{kk'} &= \omega_{k'k}, & \theta_{kk'} &= \overline{\theta}_{k'k}, \nonumber \\
    \nu_{kk'}    &= \nu_{k'k},    & \chi_{kk'}   &= \overline{\chi}_{k'k}. 
\end{align}
The concentration variables are related to these quantities by
\begin{equation}
   \left\lbrace 
   \begin{split}
    & s_{k} = \omega_{kk'} + \overline{\chi}_{kk'} + \phi_{kk'}, \\
    & \rho_{k} = \psi_{kk'} + \overline{\phi}_{kk'} + \overline{\theta}_{kk'}, \\
    & r_{k} = \nu_{kk'} + \theta_{kk'} + \chi_{kk'}.
   \end{split}
   \right.
   \label{conditional}
\end{equation}

Based on equations \ref{sirk}, we can find a time evolution equation for $\phi_{kk'}$ considering its neighborhood. The time evolution equations for the remaining pairs can be made; however, it won't be necessary for our work. For $\phi_{kk'}$ we have
\begin{multline}
    \frac{d\phi_{kk'}}{dt} = -\left(\mu_{r}+ \frac{\mu_{c}}{k'}\right)\phi_{kk'}  + \mu_{c}(k'-1) \sum_{k''} \frac{[0_{k},0_{k'},1_{k''}]}{k''}\times \\ \times P(k''|k') 
    -\mu_{c}(k-1) \sum_{k''} \frac{[1_{k''},0_{k},1_{k'}]}{k''}P(k''|k).
    %+ \alpha\theta_{ij} WANNING IMUNITY
    \label{pair1}
\end{multline}

 The first term corresponds to auto annihilation of pair concentration by a cure of the infected $k'$-degree node with rate $\mu_{r}$ or an infection of $k$-degree node by its neighbor at rate $\mu_{c}/k'$. The second and third terms are the creation and annihilation of $\phi_{kk'}$ considering all possible forms of the pair to connect with infected neighbors of different degrees, excluding the link that constitutes the pair itself ($k'-1$ term). Thus, $[0_{k},0_{k'},1_{k'}]$ and $[1_{k''},0_{k},1_{k'}]$ are the triplet formation terms. The problem is considering triplet time evolution in the same way as eq. \ref{pair1}, and later quadruplets, and so on. We can truncate this for triplet and do pair approximations 
 \begin{equation} 
    [A_{k},B_{k'},C_{k''}] \approx \frac{[A_{k},B_{k'}][B_{k'},C_{k''}]}{[B_{k'}]}.
 \end{equation}
    So eq. \ref{pair1} yields 
  \begin{multline}
    \frac{d\phi_{kk'}}{dt} = -\left(\mu_{r}+ \frac{\mu_{c}}{k'}\right)\phi_{kk'}  + \mu_{c}(k'-1)\times\\ \times\sum_{k''} \frac{[0_{k},0_{k'}][0_{k'},1_{k''}]}{[0_{k'}]k''} P(k''|k') +\\ - \mu_{c}(k-1)\sum_{k''} \frac{[1_{k''},0_{k}][0_{k},1_{k'}]}{k''[0_{k}]}P(k''|k),
    \label{phikk}
    %\alpha\theta_{kk'} wanning imunity term
\end{multline}

and by identifying pair concentrations, we can write
\begin{multline}
       \frac{d\phi_{kk'}}{dt} = -\left(\mu_{r}+ \frac{\mu_{c}}{k'}\right)\phi_{kk'} + \mu_{c}(k'-1) 
 \frac{\omega_{kk'}}{s_{k'}}\sum_{k''} \frac{\phi_{k'k''}}{k''} \times\\ \times P(k''|k') -  \mu_{c}(k-1)\frac{\phi_{kk'}}{s_{k}}\sum_{k''} \frac{\phi_{kk''}}{k''}P(k''|k).
 % wanning imunity + \alpha\theta_{kk'} 
 \label{phi-time}
\end{multline}

  To find an epidemic threshold based on linear stability around $\rho_{k}=0 $, it is necessary to adopt the quasi-static approximation $d\phi_{kk'}/dt \approx0$, $d\rho_{k}/dt \approx 0$ and discard pair-pair interaction terms. Nearly the transition point, the network is almost full of susceptible individuals; thus $\omega_{kk'} = [0_{k},0_{k'}] \approx s_{k} \approx 1-h \approx 1 $, and eq. \ref{phi-time} yields
\begin{multline}
    \left(\mu_{r}+ \frac{\mu_{c}}{k'}\right)\phi_{kk'} = \mu_{c}(k'-1) 
\sum_{k''} \frac{\phi_{k'k''}}{k''}P(k''|k').%+ \alpha\theta_{kk'}.
\label{quasiphikk}
\end{multline}

%And for 
%\begin{equation}
%     \mu\rho_{k} = \mu_{c} k \sum_{k'} \frac{P(k'|k)}{k'}\phi_{kk'}
%\end{equation}

It's easy to see using \ref{sirk} that eq. \ref{quasiphikk} is simplified to
\begin{equation}
    \left(\mu_{r}+ \frac{\mu_{c}}{k'}\right)\phi_{kk'} = \mu_{r}\rho_{k'} -\rho_{k'}\frac{\mu_{r}}{k'},
    %\alpha\theta_{kk'}  wanning imunity term
\end{equation}
\begin{comment}
From eq. \ref{finaltheta}, we have
\begin{equation}
    \theta_{kk'} = \frac{\mu}{(\alpha+2\mu)}(\rho_{k'}-\phi_{kk'})
\end{equation}
And substituting on last one
\begin{equation}
     \left(\mu_{r}+ \frac{\mu_{c}}{k'} + \frac{\alpha\mu}{\alpha + 2\mu}\right)\phi_{kk'} = \left(\frac{\alpha\mu}{\alpha + 2\mu} + \mu\left( 1-\frac{1}{k'}\right)\right)\rho_{k'}
                                                \end{equation}

For CP standard model $\alpha \rightarrow \infty$ and $\mu=1$ (no waning immunity and unitary recovery rate). Taking the limit 
\begin{equation}
    \left(2 + \frac{\mu_{c}}{k'}  \right)\phi_{kk'} = \left(2  -\frac{1}{k'}\right)\rho_{k'}
\end{equation}
Giving
\begin{equation}
    \phi_{kk'} = \frac{2k'-1}{2k'+\mu_{c}}\rho_{k'}
\end{equation}
For SIR-CP model, $\alpha \rightarrow 0$ \end{equation}
\begin{equation}
    \left(\mu_{r}+ \frac{\mu_{c}}{k'}  \right)\phi_{kk'} = \mu\left( 1 -\frac{1}{k'}\right)\rho_{k'}
\end{equation}
\end{comment}
and, finally, yielding
\begin{equation}
       \phi_{kk'} = \frac{\mu_{r}(k' -1)}{k'\mu_{r}+ \mu_{c}}\rho_{k'}.
\end{equation}
%This is the expression that we need to comeback to \ref{refcp}. 
%\begin{equation}
%      \frac{d\rho_{k}}{dt} = -\mu\rho_{k} + \mu_{c} k \sum_{k'} \frac{P(k'|k)}{k'} \frac{\mu_{r}(k' -1)}{k'\mu_{r}+ \mu_{c}}\rho_{k'}.
%\end{equation}
Therefore eq. \ref{sirk} is linearized 
\begin{equation}
    \frac{d\boldsymbol{\rho}}{dt} = \boldsymbol{\hat{J}}\boldsymbol{\rho},%\sum_{k'} J_{kk'}\rho_{k'},
    \label{linearized}
\end{equation}
where $\boldsymbol{\rho} = (\rho_{1},\rho_{2},\cdots,\rho_{k},\cdots)$ and $\boldsymbol{\hat{J}}$ is the Jacobian  matrix  which entries is given by $[J_{kk'}]$
\begin{equation}
    J_{kk'} = \frac{\mu_{r}\mu_{c} k(k'-1)P(k'|k)}{k'(k'\mu_{r}+\mu_{c})} -\mu_{r}\delta_{k',k}.
    \label{jacobian}
\end{equation}

 By diagonalizing the Jacobian matrix, the solution of equation \ref{linearized} is a sum of several exponentials. The highest eigenvalue corresponds to the leading term of this solution, and the epidemic threshold is found when it vanishes, given critical rates of infection and cure: above this value, the epidemic grows exponentially, and below it, it dies quickly. %\textcolor{red}{ Thus the solution of last equation is stable when the largest eigenvalue of this Jacobian is zero (Perron-Frobenius theorem).}
 This zero eigenvalue has an associated largest eigenvector (LEV), and, for this case, we use for the LEV\cite{barrat2008}
 \begin{equation}
    v_{k} \sim k.
    \label{LEV}
 \end{equation}

 For uncorrelated networks, $P(k'|k)$ is written in eq. \ref{conditionalprobability}, and by substituting eqs. \ref{LEV} and \ref{conditionalprobability} in the stability condition \ref{jacobian}, we obtain the following transcendental equation for cure and infection rates $\mu_{r}$ and $\mu_{c}$
\begin{equation}
    \sum_{k'} \frac{\mu_{c}\mu_{r} k'(k'-1)}{k'\mu_{r}+\mu_{c}}\frac{P(k')}{\langle k \rangle} = \mu_{r}.
    \label{transcendent}
\end{equation}

 A simple result for RRN, where $P(k) = \delta_{k,m}$ and  $m$ is the degree of all nodes, reach critical reproduction number $\lambda_{c}$
\begin{equation}
    \frac{\mu_{c}}{\mu_{r}} = \lambda^{\text{RRN}}_{c} = \frac{m}{m-2}.
    \label{localized-sir}
\end{equation}

 The result of eq. \ref{localized-sir} is the same for the Cayley tree of coordination number\cite{Tome_2011}. There is an expectation that RRN and Cayley's tree behaves as equal structure on thermodynamic limit: the former has null clustering \cite{newman-book, barabasi-review} coefficient at this stage. An identical result is obtained on \ref{phi-time} solving pair-time for earlier epidemic suppositions, neglecting pair-pair interaction and using RRN conditional probability: $P(k'|k) = k'\delta_{k',m}/m$. 
 
 For degree distributions that extend to infinity, such as those that form PLNs, equation \ref{transcendent} is a non-invertible expression to find the threshold $\lambda_{c}$. The summation is expected to converge to this solution as the largest degree $k_{c}=N^{1/\omega}$ diverges to the thermodynamic limit. Therefore, this convergence to $\lambda_{c}$ implies a threshold depending on size $\lambda_{c}(N)$. To obtain a reasonable result, we can exchange the sum of eq. \ref{transcendent} for an integral using continuous approximations for $P(k)$ and any $l$-th order moment $\langle k^{l} \rangle$, given by \cite{alencar2023droplet}, 
 \begin{equation}
    P(k) = (\gamma-1)m^{\gamma-1}k^{-\gamma}/f(\gamma -1),
  \label{distrib-pln}
\end{equation}
and 
 \begin{equation}
     \langle k^{l} \rangle = \frac{(\gamma-1)}{(\gamma-1-l)}\frac{f(\gamma-1)}{ f(\gamma-1-l)}m^{l},
 \end{equation}
 respectively, where $f(x) = 1 -(m/k_{c})^{x}$. Inserting \ref{distrib-pln} on \ref{transcendent} and integrating from $m$ to $k_{c}$ we obtain
 
 \begin{multline}
     \frac{ (\gamma-1)m^{\gamma-1}}{\langle k \rangle f(\gamma-1) } \lambda_{c}\left[\int_{m}^{\infty}\frac{k^{2-\gamma}}{k +\lambda_{c}}dk-\int_{k_{c}}^{\infty}\frac{k^{2-\gamma}}{k +\lambda_{c}}dk \right. \\
     -\left.\int_{m}^{\infty}\frac{k^{1-\gamma}}{k +\lambda_{c}}dk +\int_{k_{c}}^{\infty}\frac{k^{1-\gamma}}{k +\lambda_{c}}dk \right] = 1.
     \label{integral}
 \end{multline}

 Last expression is transformed to  Gauss hypergeometric function $F(a,b,c;z)$\cite{gradshteyn2014table}
 \begin{multline}
     \frac{ (\gamma-1)m^{\gamma-1}}{\langle k \rangle f(\gamma-1)  } \lambda_{c}\left[\frac{\Gamma(\gamma-2) m^{2-\gamma}}{\Gamma(\gamma-1)} [F(1,\gamma-2,\gamma-1;-\lambda_{c}/m)\right.\\
     \left.-\left(\frac{k_{c}}{m}\right)^{2-\gamma}F(1,\gamma-2,\gamma-1;-\lambda_{c}/k_{c})]\right.\\
     -\left.\frac{\Gamma(\gamma-1) m^{1-\gamma}}{\Gamma(\gamma)}[F(1,\gamma-1,\gamma;-\lambda_{c}/m)\right. \\
     -\left.\left(\frac{k_{c}}{m}\right)^{1-\gamma}F(1,\gamma-1,\gamma;-\lambda_{c}/k_{c})]\right] = 1.
     \label{hypergeometric}
 \end{multline}
 
  Where $\Gamma(x)$ is the usual gamma function. Here, we can analyze the complete graph limit when $m\gg1$. Expansion to low orders in $z$ in last is made on power series \cite{gradshteyn2014table}
 \begin{equation}
     F(a,b,c;z) =1 + \frac{ab}{c}z + \frac{a(a+1)b(b+1)}{c(c+1)}\frac{z^{2}}{2!} + \cdots .
     \label{hyper-expansion}
 \end{equation}

Applying eq. \ref{hyper-expansion} to \ref{hypergeometric} we have, truncating on second order or $\lambda_{c}$
\begin{comment}
    \begin{multline}
     \frac{ (\gamma-1)m^{\gamma-1}}{\langle k \rangle f(\gamma-1)  } \lambda_{c}\left[\frac{\Gamma(\gamma-2) m^{2-\gamma}}{\Gamma(\gamma-1)}(f(\gamma-2)-\frac{(\gamma-2)}{\gamma-1}\frac{\lambda_{c}}{m}f(\gamma-1))\right.\\
     -\left.\frac{\Gamma(\gamma-1) m^{1-\gamma}}{\Gamma(\gamma)}(f(\gamma-1)-\frac{\gamma-1}{\gamma}\left(\frac{\lambda_{c}}{m}\right)f(\gamma)) + \mathcal{O}(\lambda_{c}^{3})\right] = 1.
     \label{non-linear}
\end{multline}
\end{comment}    
% \begin{multline}
%     \frac{\lambda_{c}}{\langle k \rangle} \left[\langle k \rangle -1  - \lambda_{c}\left(1-\left\langle \frac{1}{k}\right\rangle\right) +  \lambda^{2}_{c}\left(\left\langle \frac{1}{k}\right\rangle-\left\langle \frac{1}{k^{2}}\right\rangle\right)\right] \approx 1.
%     \label{non-linear}
% \end{multline}
\begin{multline}
    \frac{\lambda_{c}}{\langle k \rangle} \left[\langle k \rangle -1  - \lambda_{c}\left(1-\left\langle \frac{1}{k}\right\rangle\right) + \mathcal{O}(\lambda^{2}_{c})\right] \approx 1.
    \label{non-linear}
\end{multline}
 Considering linear terms, we've finally 
\begin{equation}
     \lambda_{c} \approx \frac{\langle k \rangle}{\langle k \rangle-1}.
     \label{last}
\end{equation}
  
 Thus, the first-order approximation of the transcendent equation leads to the threshold predicted by HMF, which might be interpreted as a partial solution. A Taylor expansion to the first order around large connectivity $\langle k \rangle \gg 1$ turns eq. \ref{last} on
\begin{equation}
    \lambda_{c}(N) \approx 1 + \frac{(\gamma-2)}{(\gamma-1)m}f(\gamma-1)f(\gamma-2)^{-1}.
\end{equation}

Again, an expansion on the last term of the last equation to the second order gives dominant term scaling as
\begin{multline}
        \lambda_{c}(N) \approx 1 + \frac{(\gamma-2)}{(\gamma-1)m}+\\
        +\frac{(\gamma-2)}{(\gamma-1)m^{3-\gamma}}N^{\frac{2-\gamma}{\omega}}\left(1+m^{\gamma-2}N^{\frac{2-\gamma}{\omega}}\right).
\end{multline}

Thus, we can approximate the threshold equation in the same way as \cite{daMata_2014} for network size dependence
\begin{equation}
    \lambda_{c}(N) = \lambda_{c} + \zeta_{1}N^{-\epsilon_{1}}(1+\zeta_{2}N^{-\epsilon_{2}}).
    \label{fitting-eq}
\end{equation}
with fixed $\epsilon_{1} = \epsilon_{2} = (\gamma-2)/\omega $ and $\zeta_{1,2}$ freely.
This last expression will be important to extrapolate and obtain the estimated thresholds PLN. %In the appendix \ref{appendix}, the same result is obtained when eq. \ref{transcendent} is integrated and expanded to dominant moments to first order.   

\section{Results and discussion} 
\label{results}
We performed simulations of the asynchronous SIR model on PLNs and RRNs. We generate the networks by using the uncorrelated configuration model (UCM)\cite{catanzaro2005generation}, which means each node receives a degree $k$ based on the desired probability distribution $P(k)$, thus forming a set of nodes with edges emerging from each one or ``stubs". For PLNs, $P(k)$ is bounded on an interval ranging from minimal $k=m$ to maximum degree, scaling with its size $N$, $k_{c}=N^{1/\omega}$, to avoid structural correlations \cite{boguna2004cut}. In all considered cases, we choose $\omega=2$. For RRNs, every node receives the same degree $k = m$. After the degree selection, we randomly match the stubs, neglecting self and double connections.

\begin{figure}[!t]
  \flushleft
  \includegraphics[width = \linewidth]{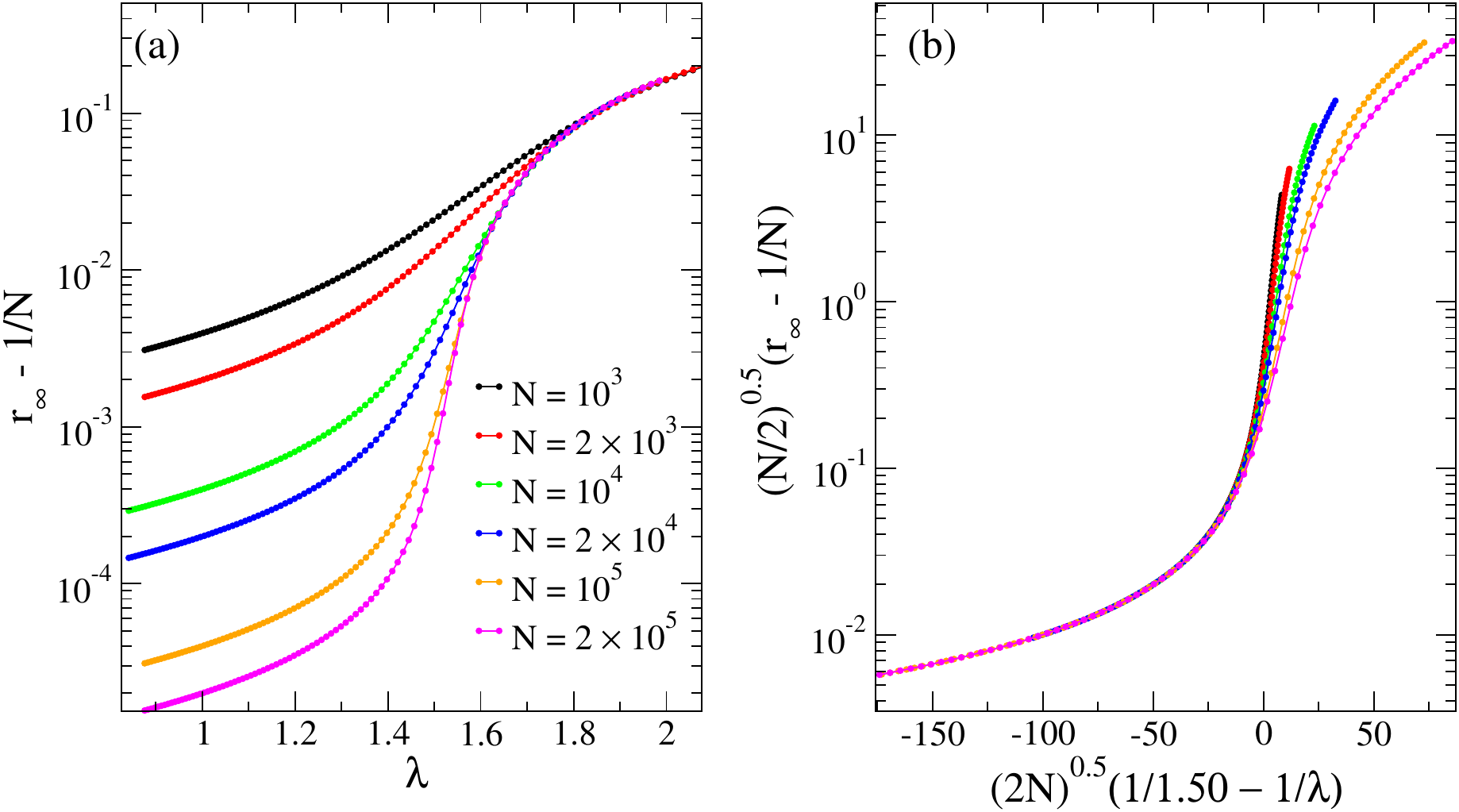}
  \caption{(Color online) Order parameter versus reproduction rate $\lambda$ (a) and its data collapse by using mean-field critical exponents (b).}
  \label{rrn-collapse}
\end{figure}

 The model was coupled to these networks using an infected list update \cite{tome2010critical, ALENCAR2020} to simulations through the Monte Carlo method. The epidemic starts from a single infected seed on an aleatory node. The dynamics occur by picking a random infected node from this list for each \textcolor{black}{time step}. With rate $\mu_{r}$, it cures spontaneously, becoming removed or recovered and erased from the list. With complementary probability, $\mu_{c} = 1-\mu_{r}$, it selects a random neighbor node, and if this last is susceptible, infect it and add to the list; otherwise, do nothing. The update continues until the network reaches an absorbing state containing only susceptible and recovered nodes. Observables  of this epidemic growth are  related to percolation \cite{newman2001fast, christensen2005complexity} $j$-moments of removed nodes
 \begin{equation}
     c^{j}_{r} = \frac{1}{N_{r}}\sum_{s} s^{j+1}n_{\text{cluster}}(s,\lambda),
 \end{equation}
 where $N_{r}$ is the number of removed nodes and $n_{cluster}(s)$ is the $s$-size cluster density of removed nodes. In this model, there is only one cluster grown for every dynamic, $n_{cluster}(s,\lambda) = \delta_{s,N{r}}$. The order parameter, fraction of removed nodes, is defined as
 \begin{equation}
     r_{\infty}(\lambda,N) = \frac{\left[ \langle c^{1}_{r}\rangle\right]}{N}.
 \end{equation}
 The symbols $[\cdots]$ and $\langle \cdots \rangle$ denote averaging on quenched realizations of networks and sample average at absorbing configuration, respectively. Recently, we proposed a cumulant to estimate critical based on second and third-order cluster moments \cite{alencar2022epidemic}
 \begin{equation}
     U = \frac{\left[ \langle c^{2}_{r}\rangle\right]^{2}}{\left[ \langle c^{1}_{r}\rangle\right]\left[ \langle c^{3}_{r}\rangle\right]}.
     \label{cumulant}
 \end{equation}
 
 The usual cumulant defined for epidemic growth in regular lattices \cite{deSouza-2011} cannot be used here because of unbound geometry complex networks. First, we test scaling relations with network size $N$ dependence from previous eq.\ref{fss-sir} nearly transition point $\lambda_{c}$ for order parameter observable on RRNs. 
 \begin{comment}
 \begin{equation}
     r_{\infty} - h \approx \left(  \frac{gN}{2a}\right)^{-1/2}F_{N}\left[\left(   \frac{1}{\lambda_{c}}- \frac{1}{\lambda} \right) \left(  \frac{2N}{ga}\right)^{1/2}\right],
\end{equation}

and cumulant \textcolor{red}{decide if it'll be used or not}
\begin{equation}
    U \approx F_{N}\left[\left(   \frac{1}{\lambda_{c}}- \frac{1}{\lambda} \right) \left(  \frac{2N}{ga}\right)^{1/2}\right]
\end{equation}
 \end{comment}
In Figure \ref{rrn-collapse}, we show order parameter curves as functions of reproduction rate and their collapses using mean-field exponents for RRNs with $g$ constant and $m=6$. The estimated threshold was $\lambda_{c} = 1.500$. Each point for each curve is an average with at least $10^{5}$ growth clusters and $400$ quenched disorders.

\begin{figure}[!t]
  \centering
  \includegraphics[scale=0.35]{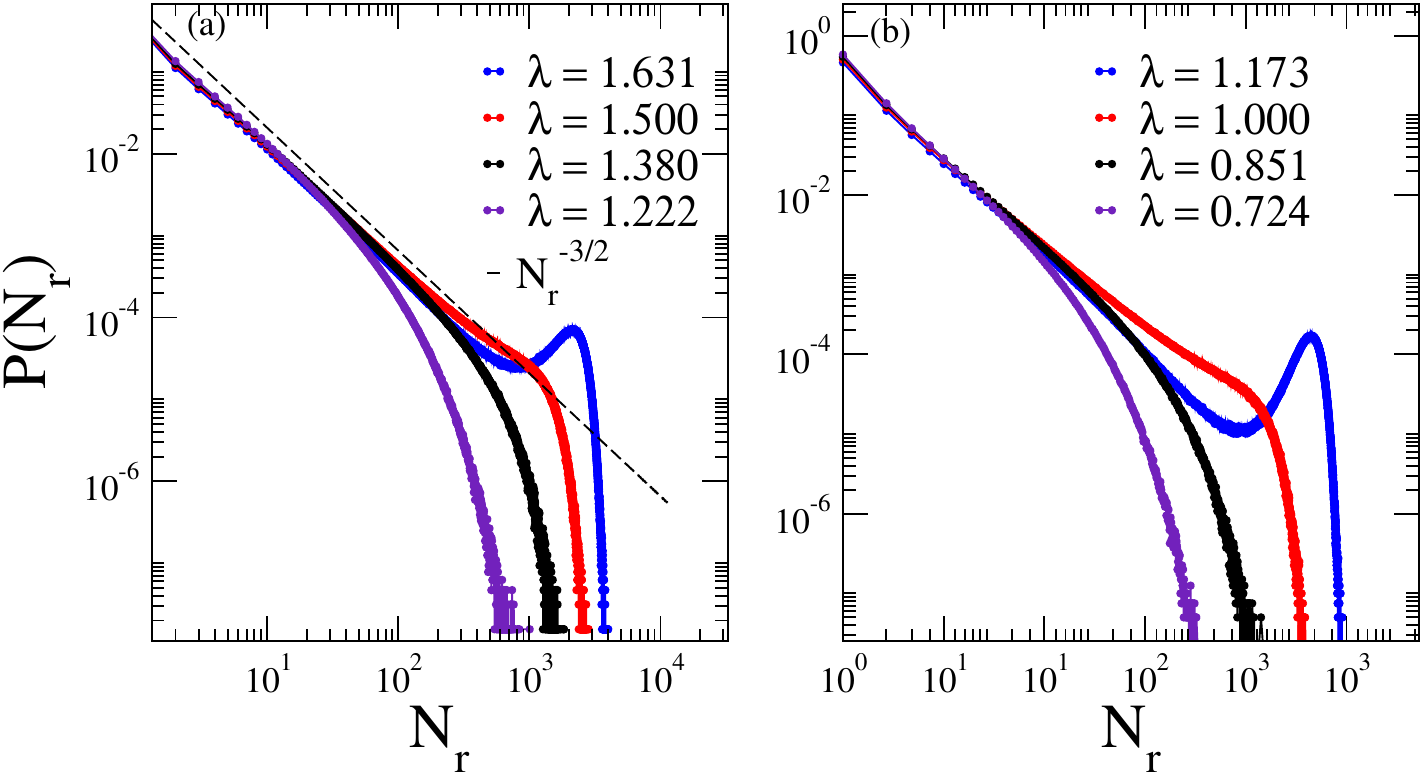}
  \caption{(Color Online) (a) Outbreak distribution for RRN with $m=6$ and $10^{4}$ nodes at different rates. The dashed line represents expected decay for homogeneous networks at infinity size and transition point $\lambda_{c}$. We did quenched averages with at least 600 replicas. (b) Outbreak distributions for annealed network using $\gamma=2.50$, $m=6$, $N=10^{4}$, and $k_{c} = N^{1/2}$.}
  \label{distribuicao}
\end{figure}

There is a good agreement on the endemic phase (below transition point), highlighting the importance of droplet term $h \approx 1/N$ at this phase. But when the system is slightly below and above the threshold, all curves tend to separate from collapse. This result is quantitatively comparable with previous ones for PLN \cite{alencar2022epidemic} and annealed graphs (non-shown data), all these showing the same behavior. Therefore, this effect is not due to the presence or absence of connections fixed on time but to the intrinsic dynamics of the model itself. Reference \cite{ben2004size} arguments in favor of a non-trivial mean-field exponent for average epidemic size $\langle n\rangle$, which scales as $\sim N^{1/3}$ for complete graphs at transition point $\lambda_{c}=1$.

 Some authors \cite{kessler-2007, shu2015numerical} point to a bimodal distribution on the standard SIR model on complete graphs and RRNs for the supercritical phase. In Figure \ref{distribuicao}, we show in the asynchronous SIR epidemic outbreaks distribution $P(N_{r})$ for RRNs and annealed PLN for three regimes: sub-critical, critical, and supercritical. In figure \ref{distribuicao}-(a) for the sub-critical and critical phases, the epidemic distribution seems to show a power-law decay, agreeing with theoretical result $P \sim N_{r}^{-3/2}$ \cite{Tome_2011}, until a maximum before a decaying.

 The power-law decay explains a good collapse on Fig. \ref{rrn-collapse} for this region: the average epidemic size per site scales as $\sim N^{-1/2}$ (one can see this integrating $N_{r}P(N_{r})$ from 1 to $N$ and dividing by $N)$. After the power-law behavior, there is a fast decay, which is negligible for calculating moments and expected to shift to zero at infinity network size. For the super-critical phase, the rapid decline gives place to a peak, changing the epidemic distribution to a bimodal form. Significantly, the average epidemic size per site (order parameter), expected as $~N^{-1/2}$, changes for this region. In Figure \ref{distribuicao}-(b), we show the distribution of the outbreak for annealed networks to confirm non-structural characteristics of bimodal behavior. 

\begin{figure}[!t]
     \centering
   \includegraphics[width=\linewidth]{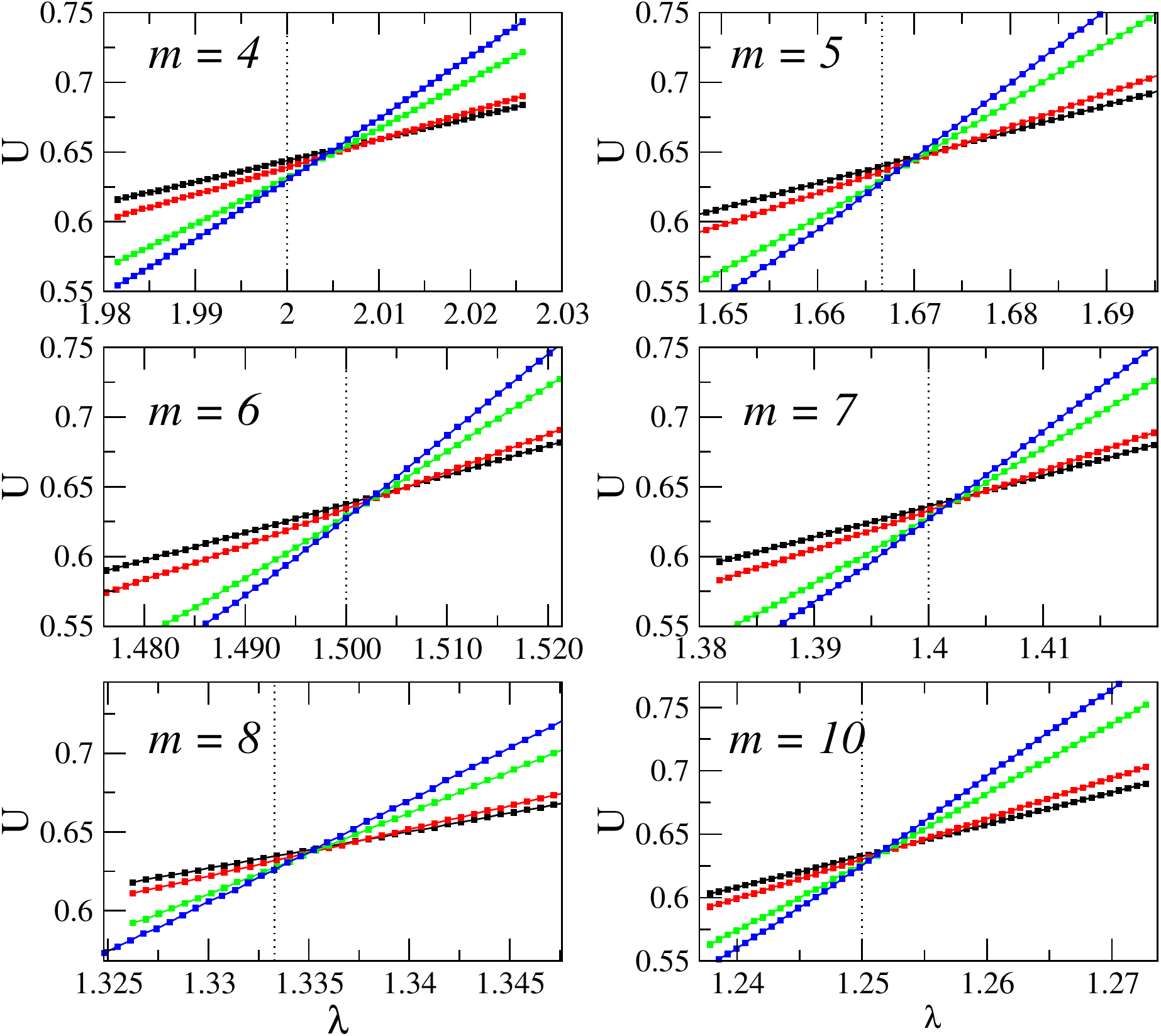}        
     \caption{(Color online) Third-order cumulant for RRNs. Network sizes used for each panel are $N = 10^{4}$, $2\times 10^{4}$, $10^{5}$, and $2\times 10^{5}$ nodes (black, red, green, and blue points, respectively). Crossings between two consecutive curves give an estimate of the transition point. Dashed vertical lines are theoretical predictions by eq. \ref{localized-sir}.
     %Suggestion: put RRN m=8 and ER on same graph but shifting one of them to up to see the same transition point
     }
     \label{rrn-cumulant}
\end{figure}

However, the peak presence does not affect the transition point estimation by eq. \ref{cumulant}. Thresholds for RRNs with varying connectivity $m$ were estimated using cumulant given in eq. \ref{cumulant}. In Figure \ref{rrn-cumulant}, we show this quantity using the same simulation parameters of Fig. \ref{rrn-collapse}. Table \ref{tabela-rrn} summarizes critical thresholds obtained by simulations and theoretical prediction, eq. \ref{localized-sir}, values of $m$ used for RRN. The thresholds measured strongly agree with the theory, almost exactly when $m \gg 1$. 

 %\lipsum[1-1]

  %For PLN with $m = 8$, \textcolor{red}{ is necessary to run larger sizes for UCM networks to see a clearly estimated transition point to discard overestimating of PHMF theory}

We estimated the thresholds for PLNs by solving the transcendental equation \ref{transcendent}, and the roots are functions of the network parameters (e.g. $\gamma, m, \langle k \rangle$, $k_{c}(N)$). The critical threshold is also dependent on the network size $\lambda_{c}(N)$. We used the bisection method to solve eq. \ref{transcendent} for each network size $N$ fixing $\gamma$, $m$, $k_{c}$, and $\langle k\rangle$ given by
\begin{equation}
\langle k \rangle=\frac{\sum_{k=m}^{k_{c}}k^{1-\gamma}}{\sum_{k=m}^{k_{c}} k^{-\gamma}},
\end{equation}
and the degree distribution 
\begin{equation}
  P(k)=\frac{k^{-\gamma}}{\sum^{k_c}_{k=m} k^{-\gamma}}.
\end{equation}
  
\begin{figure}[!t]
    \centering
    \includegraphics[scale = 0.35]{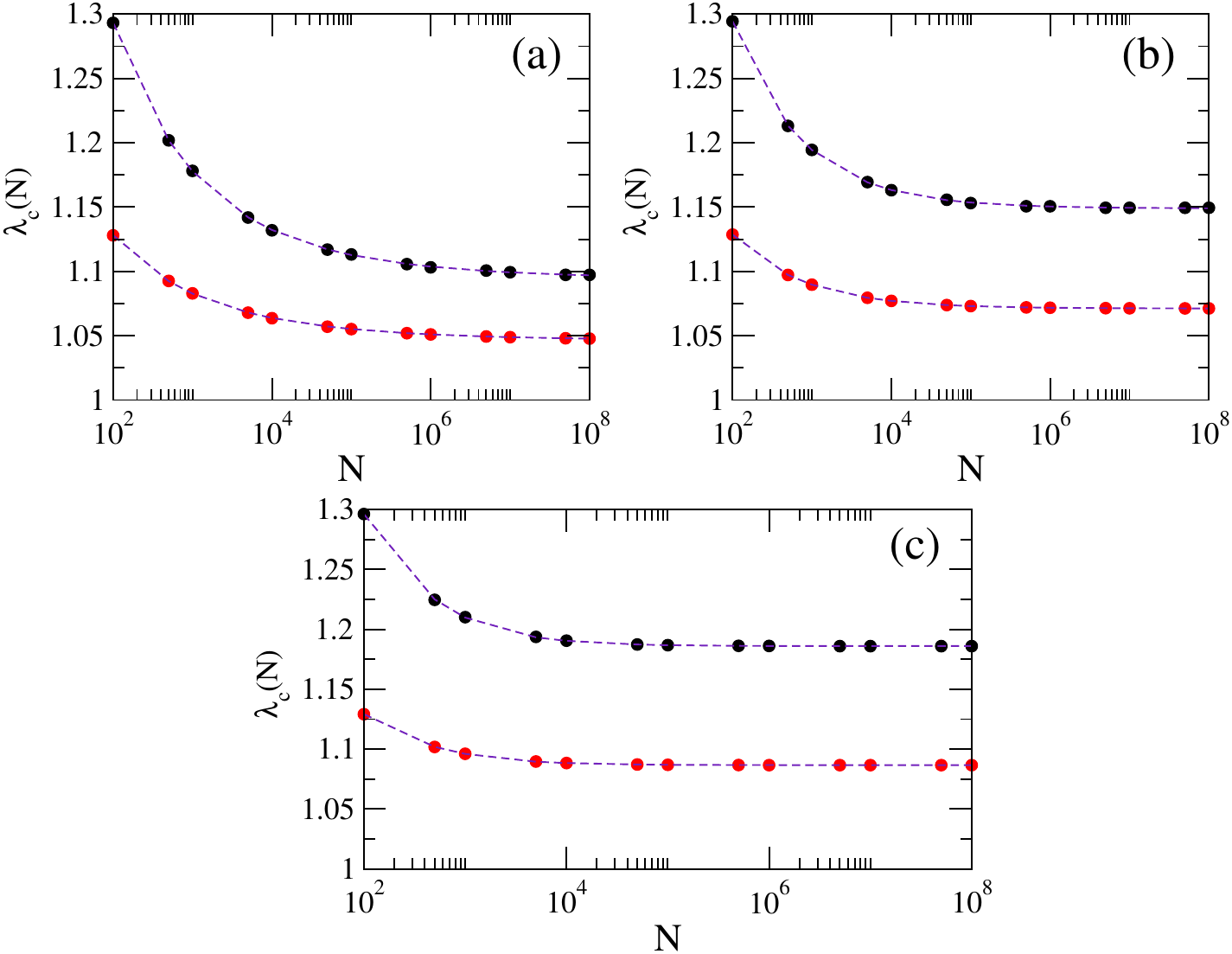}
    \caption{(Color online). Critical basic reproduction number as a function of network size $\lambda_{c}(N)$ for first-order approximation (red), eq.\ref{last}, and for solutions of transcendent equation (black) from eq. \ref{transcendent} on PLN with $m = 8$ and $\gamma = 2.50, 3.00, 3.50$ ((a), (b), and (c) panels, respectively). Dashed lines are non-linear fitting obtained by equation \ref{fitting-eq}. }
    \label{ucm-decay}
\end{figure}

  Figure \ref{ucm-decay} shows this dependence for transcendent equation solutions and first-order approximation from eq. \ref{last}. Both plateau regions indicate a slow decay of $\lambda_{c}(N)$ towards the thermodynamic limit $N \rightarrow \infty $. Interestingly, heuristic approximation and transcendent solution converge similarly, justifying using eq. \ref{last}, and, thus, \ref{fitting-eq} to extrapolate this quantity at infinity.  
  
\begin{figure}[!h]
     \centering
     \includegraphics[width=\linewidth]{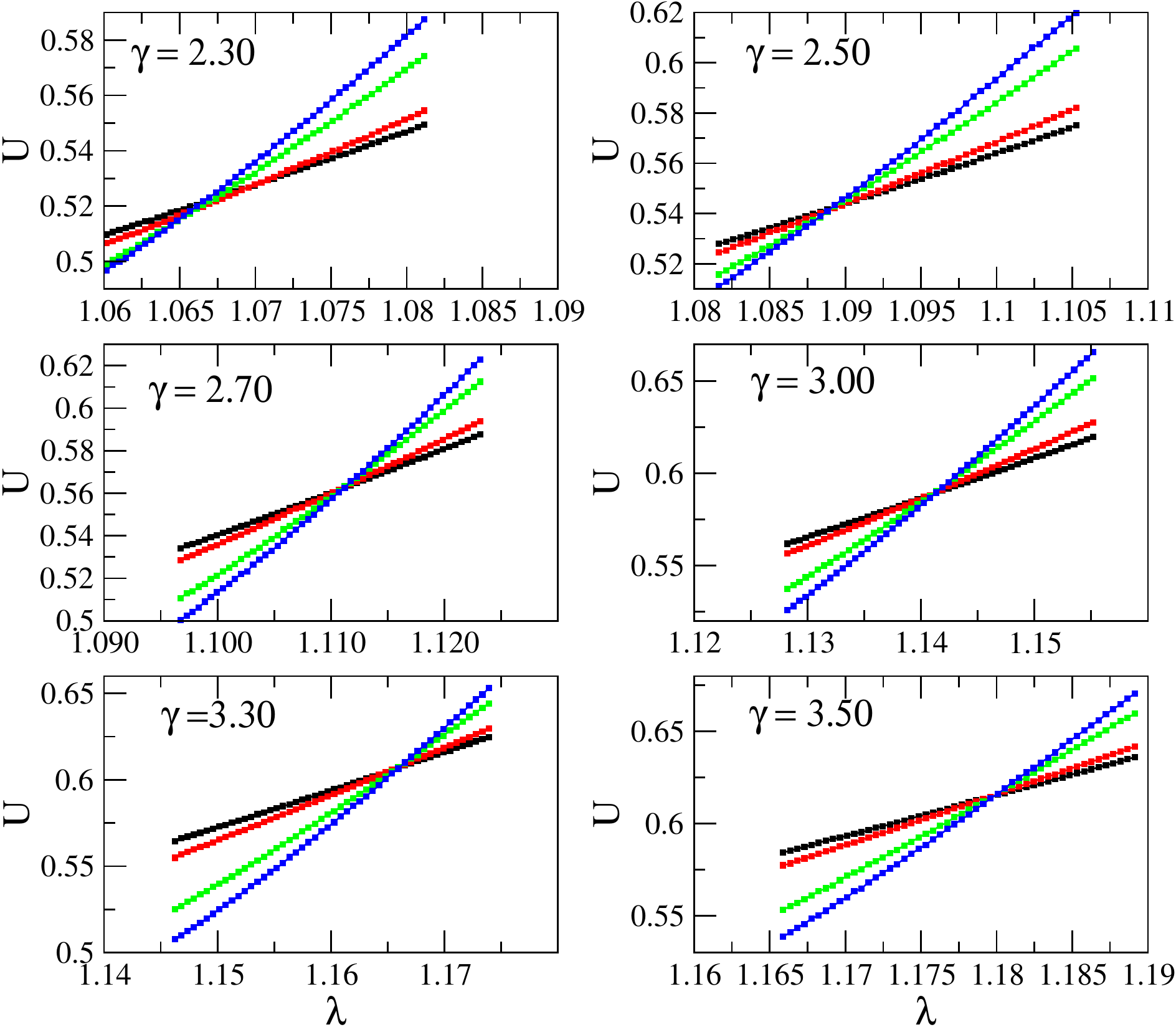}
     \caption{ (Color Online) Third order cumulant for PLN with for $m=8$. Network sizes used from curve less steep to steeper for each panel were $N = 10^{4}, 2\times 10^{4}, 10^{5}, 2\times 10^{5}$ nodes (black, red, green, and blue points, respectively).}
     \label{cumulant-ucm}
 \end{figure}

We performed the simulations to compare with theoretical predictions, ranging  $\gamma$ and fixing $m=4$ (non-show) and $m=8$ for PLNs. In figure \ref{cumulant-ucm}, we show crossings for different curves using $10^{5}$ ensembles and $800$ quench disorders. Results for values of $\gamma$ for $m=4$ and $m=8$ are summarized in table \ref{tabela-pl}.
 
  %\textcolor{red}{ Check all extrapolations using  CP model as reference}
 %\lipsum[1-1]
 \begin{table}[!h]
  \begin{center}
 \begin{tabular}{  m{2em}  m{4em}  m{3.2em} } 
   \hline
   \textbf{$m$}& \textbf{$\lambda^{*}_{c}$}  &$\lambda^{\text{PHMF}}_{c}$\\ 
   \hline
   4 & $2.004(4)$ & $2.000$\\ 
   5 & $1.670(3)$ & $1.666$\\ 
   6 & $1.502(2)$& $1.500$ \\ 
   7 &$1.403(1)$ & $1.400$\\ 
   8 & $1.335(1)$&$1.333$\\ 
   10 &$1.251(2)$& $1.250$\\ 
   \hline
 \end{tabular}
 \caption{Critical thresholds $\lambda^{*}_{c}$ on simulations compared with theoretical with PHMF $\lambda^{\text{PHMF}}_{c}$. Numbers in parenthesis are uncertain in the last decimal places.}
 \label{tabela-rrn}
 \end{center}
 \end{table}
 
 \begin{table}[!h]
 \begin{center}
  \begin{tabular}{ m{2em} | m{4em} m{3.2em} m{3em} | m{4em} m{3em} m{3em} }
  \hline
  & \multicolumn{3}{c|}{m=8}  &  \multicolumn{3}{c}{m=4} \\
  \hline
  \textbf{$\gamma$}& $\lambda^{*}_{c}$  &$\lambda^{\text{PHMF}}_{c}$&$\lambda^{\text{HMF}}_{c}$ & $\lambda^{*}_{c}$  &$\lambda^{\text{PLN}}_{c}$ &$\lambda^{\text{HMF}}_{c}$\\ 
  \hline
  2.30 & $1.066(6)$& $1.067$& $1.029$&$1.134(8)$&$1.145$&1$.061$ \\ 
  2.50 & $ 1.088(1)$& $1.093$&  $1.043$&$1.189(3)$&  $1.212$&$ 1.090$\\ 
  2.70 &$1.110(1)$ &$1.118$&$1.054$& $1.247(4)$&$1.277$&$1.114$\\ 
  3.00 & $1.140(1)$ &$1.149$&$1.066$& $1.329$(3)&$1.364$&$1.142$ \\ 
  3.30& $1.165(2)$&$1.172$&$1.076$&$1.402(2)$&$1.437$&$1.164$\\ 
  3.50 & $1.179(1)$& $1.186$&$1.081$& $1.445(1)$&$1.480$&$1.176$ \\ 
  \hline
\end{tabular}
\caption{\label{demo-table} Simulated critical thresholds $\lambda^{*}_{c}$ vs. thresholds obtained by PHMF theory $\lambda^{\text{PHMF}}_{c}$ vs. thresholds obtained by HMF theory $\lambda^{\text{HMF}}_{c}$ for PLNs. Parenthesis indicates numerical uncertainty from cumulant crossings on the last digit.}
\label{tabela-pl}
\end{center}
\end{table}
%\textcolor{blue}{talk about overestimation on transition point}

\begin{comment}
\begin{table}
\caption {\label{tab:table1} Estimated critical screening parameters of Hulth\'en potential, for 
some high-lying states having $n=6-10, \ell=0-9$, along with literature results.} 
\begin{ruledtabular}
\begin{tabular}{llllll}
 $\gamma$ & \multicolumn{2}{c}{$m$} & State & \multicolumn{2}{c}{$\delta_c$}    \\ 
\cline{2-3}  \cline{5-6} 
        &  $4$   & $8$ 
        \\ \hline %   &             
	%&  PR$^\dag$   & Literature                                                  \\    \hline
 $6s$   &  0.0555555   & 0.055556\footnotemark[1]                                    & \\ 
 %$6g$   & 0.04058464   & 0.040585\footnotemark[1],0.04058464\footnotemark[2]         \\       
 $7s$   &  0.0408163   & 0.040816\footnotemark[1]                                    & 
        
                                  \\
\end{tabular}
\end{ruledtabular}
\begin{tabbing}
$^{\mathrm{a}}$Ref.~\cite{varshni90}. \hspace{25pt} \= 
$^{\mathrm{b}}$Ref.~\cite{demiralp05}. \hspace{25pt} \=
$^\dag$PR implies Present Result.
\end{tabbing}
\end{table}

\begin{tabular}{llllr}
\firsthline
\multicolumn{2}{c}{Item}  \\
\cline{2-3} 7  
$\gamma$    & $\lambda_{c}^{*}$ &$\lambda_{c}^{UCM}$ & a& a\\
\hline
Gnat      & per gram    & 13.65      \\
          & each        & 0.01       \\
Gnu       & stuffed     & 92.50      \\
Emu       & stuffed     & 33.33      \\
Armadillo & frozen      & 8.99       \\
\lasthline
\end{tabular}
\end{comment}

For a last case, we apply equation \ref{transcendent} on preferential linear attachment Barabási-Albert (BA) networks \cite{barabasi1999} to compare with previous simulations on the asynchronous SIR model \cite{alencar2023droplet}. The BA network degree distribution $P(k)\sim k^{-3}$ classifies it as marginal scale-free (second moment divergent logarithmically with size $N$)\cite{barabasi-review}. Because of its growth process, intuitively, one might expect considerable correlations expressed as assortative or disassortative mixing in this network model\cite{barrat2008, newman-book}. However, it was shown that it can be considered almost neutral \cite{RPS-2005}, and HMF can be applied successfully to describe scaling non-equilibrium models \cite{alencar2023droplet}. Data from crossings and theoretical results were listed in table \ref{tabela-ba} for BA networks varying minimal connectivity $m$.
%\textcolor{red}{Decide if will concatenates BA data vs m and UCM vs. gamma on only graph}

\begin{figure}[!t]
    \centering
\includegraphics[width=\linewidth]{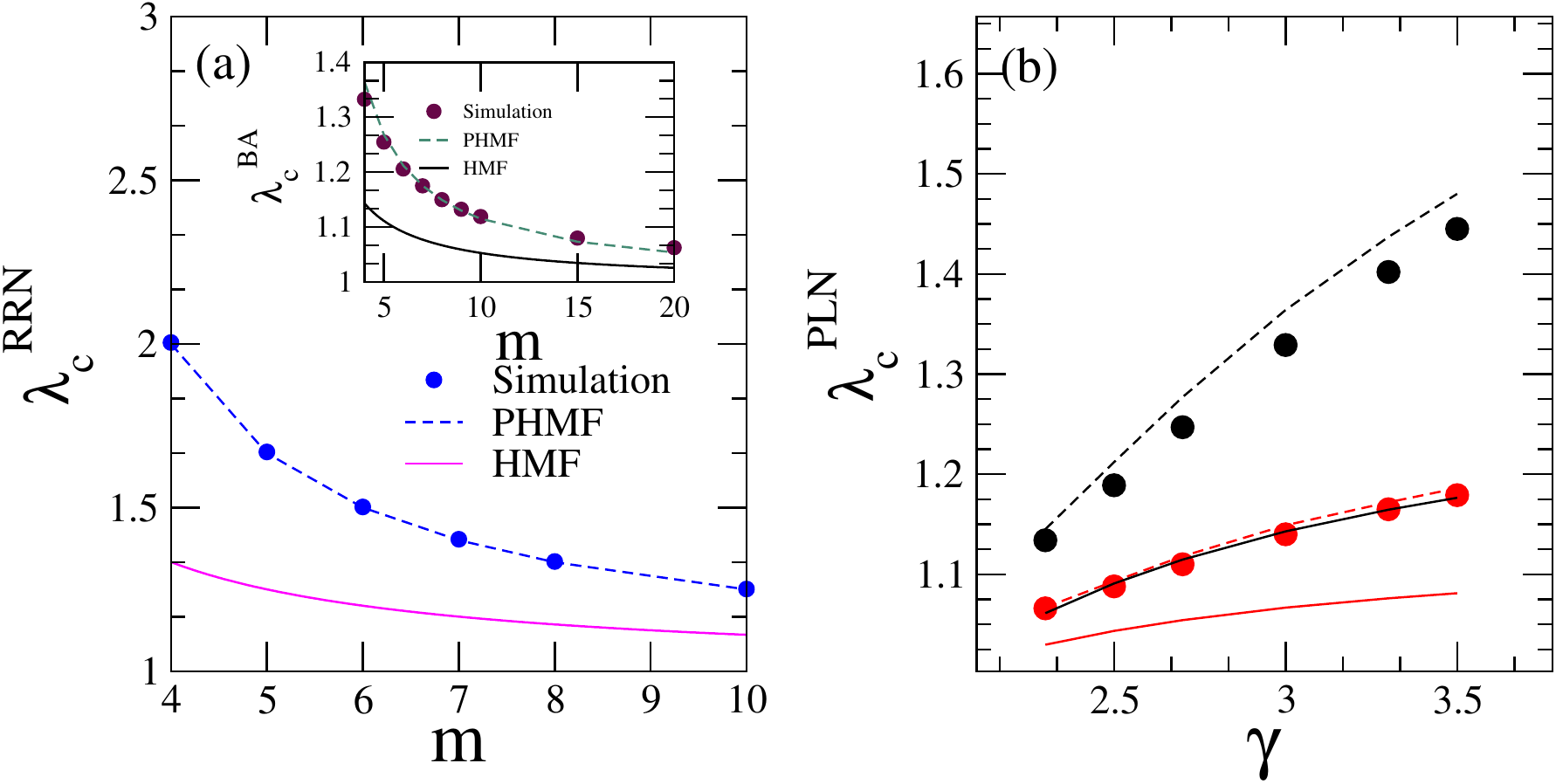}
    \caption{(Color online) We show a phase diagram where the curves show the epidemic thresholds (dots) as functions of the degree in (a) for RRNs, and of the exponent distribution in (b) for PLNs with $m=4$ (black dots) and $m=8$ (red dots). In both cases, we compare the thresholds obtained by HMF (solid lines) and PHMF (dashed lines) approximations with simulations. We show simulation points of the thresholds for PLN for . Inset: results for BA networks as a function of $m$.}
    \label{phase-space}
\end{figure}

Figure \ref{phase-space} summarizes our results for the three networks studied. For RRN type, PHMF offers a better improvement about HMF,  as is expected, on showed simulations, being almost exact when $m \gg 1$. This behavior indicates one of two recipes to mean-field accuracy: more homogeneity. In the PLN type, otherwise, there is a high sub estimation for HMF and low overestimation for PHMF in all conditions simulated. The case with $m=8$ gives a better result than $ m=4$ when $\gamma \rightarrow 2 $. In this case, we believe that by holding $\gamma$ and increasing $m$, the average distances for the same PLN sizes $N$ is shorter. Therefore, participation in a larger $k-$degree node on epidemic spreading is more possible, agreeing more with eq. \ref{transcendent}. This is the second item to better accuracy: less distances. Results for the BA network went from overestimated to underestimated when we tuned $m$; thus, this corroborates the accuracy mean-field hypothesis and the low correlation expected from the network.  Sizes used here $N=4900$ to $N=14400$
    %%%%%%%%%%%%%%%%%%%%%%%%%%
    %\textcolor{red}{At the same time the clusterization coefficient participates of the process. For $ 2<\gamma\leq 3$ the local clustering is $C \sim N^{2-\gamma}$, vanish with $N \rightarrow \infty$, but number of triangles is diverging. For $\gamma > 3$, $C \sim N^{-1}$ but number of triangles remain finite. This can be the source (\textcolor{green}{ this is a speculation}) of overestimation: high triangles can produce an effective low impact on simulations while low ones produces significantly deviations of theory. } 
%%%%%%%%%%%%%%%%%%%%%%%%%%%%%%%%%%%%%%%%%%%

\begin{table}[!t]
  \begin{center}
 \begin{tabular}{ m{2em}  m{3em} m{3em} m{3em} } 
   \hline
   $m$& $\lambda^{*}_{c}$ &$\lambda^{\text{BA}}_{c}$&$\lambda^{\text{HMF}}_{c}$ \\ 
   \hline
     4 & $1.332$ & $ 1.364$&$1.142$\\ 
    5 & $1.254$ & $ 1.268$&$1.111$\\ 
   6 & $1.205$& $1.211$&$1.090$ \\ 
    7 &$1.174$ & $1.175$&$1.076$\\ 
     8 & $1.149$&$1.148$&$1.066$\\ 
     9 &$1.132$& $1.129$&$1.058$\\ 
     10 &$1.119$& $1.114$&$1.052$\\ 
     15 &$1.079$& $1.073$&$1.034$\\ 
    20 &$1.062$& $1.053$&$1.025$\\ 
    \hline
 \end{tabular}
 \caption{Critical thresholds $\lambda^{*}_{c}$ on simulations compared with theoretical with PHMF $\lambda^{\text{BA}}_{c}$ and HMF
 $\lambda^{\text{HMF}}_{c}$for BA networks.}
 \label{tabela-ba}
 \end{center}
 \end{table}
%\textcolor{red}{Maybe for $m=8$ there's some overestimating. Testing to $m=3$ $\cdots$. Or, slow decay with cutoff exponent it isn't possible see for large network the critical threshold plateau the infinity size thresholds, even for heuristic approximation \ref{hpa}. So the thresholds on right handed size of table 1 need to be corrected.x'x }

\section{Conclusions} 
\label{conclusion} 

  The HMF data collapse agrees with the simulation results of SIR dynamics in the subcritical phase and deviates from the simulation results in the supercritical phase. This result came from the distribution of epidemics (removed sites), having power-law distribution at the first phase and bimodal form at the later, which alters order parameter scaling. For a more precise collapse, we speculate there must be a way to find an outbreak cutoff to exclude the peak from the simulation, which constitutes a challenging task for RRN, PLN, and their annealed versions.
  
  We use a cumulant, robust on both phases, to estimate the epidemic thresholds. It indicates a reliable way to do it, particularly for RRN networks, where pair theory and simulation show a precise agreement. A minor overestimation was found in PLN networks. This result can be associated with the quenched network structure since the expression to conditional probabilities $P(k'|k)$ is exact for annealed networks, as mentioned on eq. \ref{conditionalprobability}, resulting in epidemic thresholds $\lambda_c=1$ as seen on the figure \ref{cumulant-annealed}.

    If we discard the low correlation effect for BA networks, a transition from overestimation to underestimation is evidence that PHMF for our model works as well as with more homogeneity, showing a window of convergence. We hope this work serves as an approximation to calculate epidemic thresholds for the asynchronous SIR model in real networks when their metric parameters (e.g., $P(k)$ and $\langle k \rangle$) are inserted on transcendent equation \ref{transcendent}.

\section*{Acknowledgments} \label{sec:acknowledgements}
We would like to thank CAPES (Coordenação de Aperfeiçoamento de Pessoal de Nível Superior), CNPq (Conselho Nacional de Desenvolvimento Científico e tecnológico),  and FAPEPI (Fundação de Amparo à Pesquisa do Estado do Piauí) for the financial support. We acknowledge the \textit{Dietrich Stauffer Computational Physics Lab.}, Teresina, Brazil, where all computer simulations were performed. We thank  Angélica S. Mata for her helpful insight into this work. 
     %We would like to thank to us. 
    % RENEMBER OF ACKNOWLEDGE DRA. ANGELICA DA MATTA FOR DISCUSSION«
\bibliographystyle{elsarticle-num-names}
\bibliography{textv1}

\appendix%* descomentar para tirar o rótulo do apêndice

\end{document}